\documentstyle[art12]{article}
\catcode`\@=11 \@addtoreset{equation}{section}\catcode`\@=12
\newcommand{\beq}[1]{\begin{equation}\label{#1}}
\newcommand{\eeq}{\end{equation}}
\newcommand{\bear}[1]{\begin{eqnarray}\label{#1}}
\newcommand{\ear}{\end{eqnarray}}
\newcommand{\nn}{\nonumber}
\newcommand{\rf}[1]{(\ref{#1})}

\def\noi{\noindent}

\newcommand{\Title}[1]{\noi {\Large #1} \\}
\newcommand{\email}[2]{\footnotetext[#1]{e-mail: #2}}%
\newcommand{\foom}[1]{\protect\footnotemark[#1]}
\begin{document}
\input epsf.sty
\vspace*{-2.0cm}
\centerline{\mbox{\hspace*{10cm} Preprint P-Math-98/2}}
\centerline{\mbox{\hspace*{10cm} Univ. Potsdam, 1997
}}
\vspace*{1.0cm}
\begin{center}
\Title{\large\bf
Toda chains with type $A_m$ Lie algebra for \\
multidimensional $m$-component perfect fluid cosmology}

\bigskip

\noi{\normalsize\bf
V. R. Gavrilov\foom 1\dag,
U. Kasper\foom 2\ddag,
V. N. Melnikov\foom 3\dag,
and M. Rainer\foom 4\ddag }

\medskip

\noi{\dag\ \it Centre for Gravitation and Fundamental Metrology
\\
VNIIMS, 3-1 M. Ulyanovoy St., Moscow 117313, Russia}

\medskip

\noi{\ddag\
\it Institut f\"ur Mathematik, Universit\"at Potsdam,
\\
       PF 601553, D-14415 Potsdam, Germany}
\end{center}
\bigskip

\vspace{5mm}
\begin{center}
{\bf Abstract}
\end{center}
We consider a $D$-dimensional cosmological model describing an
evolution of Ricci-flat factor spaces,
$M_1,\ldots,\\ M_n$  ($n\geq 3$),
in the presence of an $m$-component perfect fluid source ($n-1\geq m\geq 2$).
We find characteristic vectors, related to the matter constants in the
barotropic equations of state for fluid components of all factor spaces.
We show that, in the case where we can interpret these vectors
as the root vectors of a Lie algebra of Cartan type $A_m=sl(m+1,\bf{C})$,
the model reduces to the classical open $m$-body Toda chain.
Using an elegant technique by Anderson \cite{Anderson} for solving this
system, we integrate the Einstein equations for the model and
present the metric in a Kasner-like form.

PACS numbers: 04.20.J, 04.60.+n, 03.65.Ge

\email 1 {gavr@rgs.phys.msu.su}
\email 2 {ukasper@rz.uni-potsdam.de}
\email 3 {melnikov@rgs.phys.msu.su}
\email 4 {mrainer@rz.uni-potsdam.de}

\section{Introduction}
\setcounter{equation}{0}
Recently, investigations on multidimensional gravitation and cosmology
have found renewed interest.
It is well known now, that multi-scalar-tensor models
derived from a higher dimensional multidimensional Einstein action
are similar to the (bosonic sector of) effective
low-energy models from (super) string theory.
Beyond that fact, it was recently shown that the multidimensional
ansatz provides also a natural clue to membrane theory (as the natural
generalization of string theory).

Here however, we restrict to multidimensional (spacially homogeneous)
cosmology \cite{M94,M95}.
The $D$-dimensional cosmological model describing the evolution of $n$
($n\geq 3$)
Ricci-flat spaces $M_1,\ldots, M_n$ in the presence of an $m$-component
($n-1\geq m\geq 2$) perfect
fluid source is considered. The barotropic equations of state for the
mass-energy densities and pressures of the components are given for each
space. When the vectors related to the constants in the barotropic equations
of state can be interpreted as root vectors of the Lie algebra
$A_m=sl(m+1,\bf{C})$, the model reduces to the classical open-chain
$m$-body
Toda system. Using the new elegant form of its exact solution proposed in
\cite{Anderson}, we integrate the Einstein equations for the model and
present the metric in the Kasner-like form.

\section{The model and the equations of motion}
\setcounter{equation}{0}
The $D$-dimensional space-time manifold $M$  may be the product  of the
time axis $R$ and $n$ manifolds $M_1,\ldots,M_n$, i.e.
\begin{equation} 
M = R \times M_{1} \times \ldots \times M_{n}.
\end{equation}
The product of  some of the manifolds  $M_1,\ldots,M_k$, $1\leq k\leq 3$,
gives the external 3-dimensional
space and the remaining part $M_{k+1},\ldots,M_n$
stands for so-called internal spaces.
Further, for sake of generality,
we admit that dimensions $N_i={\rm dim}M_i$ for $i=1,\ldots,n$ are arbitrary.

The manifold $M$ is equipped with the metric
\begin{equation} 
{\rm d}s^2=-{\rm e}^{2\gamma(t)}{\rm d}t^2 +
\sum_{i=1}^{n}\exp[2{x^{i}}(t)]{\rm d}s^2_i,
\end{equation}
where $\gamma(t)$ is an arbitrary function determining
the time $t$ and d$s^2_i$ is the metric on the manifold $M_i$.
We assume that the manifolds $M_1,\ldots,M_n$ are Ricci-flat, i.e.
the components of the Ricci tensor for the metrics d$s^2_i$ are zero.
Under this assumption the Ricci tensor for the metric (2.2) has following
non-zero components
\begin{eqnarray} 
R_0^0= {\rm e}^{-2\gamma}\left(
\sum_{i=1}^{n} N_{i}(\dot{x}^{i})^2+  \ddot{\gamma_0}-
\dot{\gamma}\dot{\gamma_0}\right)\\
R_{n_i}^{m_i}=
{\rm e}^{-2\gamma}
\left[\ddot{x}^{i}+ \dot{x}^{i}(\dot{\gamma_0}-\dot{\gamma})\right]
\delta_{n_i}^{m_i}
\end{eqnarray}
with the definition
\begin{equation} 
\gamma_0=\sum_{i=1}^{n} N_{i}x^{i}.
\end{equation}
Indices $m_i$ and $n_i$ in (2.3), (2.4) for
$i=1,\ldots,n$ run from ($D-\sum_{j=i}^{n}N_j$) to
($D-\sum_{j=i}^{n}N_j+N_i$) ($D=1+\sum_{i=1}^{n}N_i={\rm dim}M $).

The source of gravity shall be a multicomponent
perfect fluid. The energy-momentum tensor of such source comoving
coordinates reads
\begin{eqnarray}
&& T^{M}_{N} = \sum_{s =1}^{m} T^{M
(s)}_{N}, \\
&&(T^{M (s)}_{N})={\rm diag}\left(-{\rho^{(s)}}(t),
\underbrace{p_{1}^{(s)}(t),\ldots , p_{1}^{(s)}(t)}_{N_1 times},\ldots ,
\underbrace{p_{n}^{(s)}(t),\ldots , p_{n}^{(s)}(t)}_{N_n times}\right)
\quad ,
\end{eqnarray}
Furthermore we suppose that for any $s$-th component of the perfect
fluid the barotropic equation of state is given by
\begin{equation}\label{2.8}
{p_{i}^{(s)}}(t) =\left(1-{h_{i}^{(s)}}\right){\rho^{(s)}}(t),\ \
s=1,\ldots,m,
\end{equation}
where
${h_{i}^{(s)}} = {\rm const}$.

The equation of motion
$\bigtriangledown_{M} T^{M (s)}_{0}=0$ for the $s$-th component
of the perfect fluid described by the tensor (2.7) reads
\begin{equation}
\dot{\rho}^{(s)}+
\sum_{i=1}^{n}N_{i}\dot{x}^{i}\left(\rho^{(s)} + p_{i}^{(s)}\right)=0.
\end{equation}
Using the equations of state (2.8), we obtain from (2.9) the following
integrals of motion
\begin{equation}
A^{(s)}={\rho^{(s)}}
\exp\left[2\gamma_{0} - \sum_{i=1}^{n}N_ih_i^{(s)}x^{i}\right]=
{\rm const}.
\end{equation}

The Einstein equations
$R^M_N-R\delta^M_N/2=\kappa^2T^M_N$
($\kappa^2$ is the gravitational constant), can be written as
$R^M_N=\kappa^2[T^M_N-T\delta^M_N/(D-2)]$.
Furthermore, we employ the equations
$R^{0}_{0}-R/2=\kappa^{2}T^{0}_{0}$ and
$R^{m_{i}}_{n_{i}}=\kappa^{2}[T^{m_{i}}_{n_{i}}-
T\delta^{m_{i}}_{n_{i}}/(D-2)]$. Using (2.3)-(2.8), we obtain for them
\begin{equation}
\frac{1}{2}\sum_{i,j=1}^{n}G_{ij}\dot{x}^{i}\dot{x}^{j}+ V=0,
\end{equation}
\begin{eqnarray}
\ddot{x}^{i} + \dot{x}^{i}(\dot{\gamma_0}-\dot{\gamma})&=&
- \kappa^{2} \sum_{s=1}^{m}A^{(s)}
\left(
h^{(s)}_{i} - \frac{\sum_{k=1}^{n}N_{k}h^{(s)}_{k}}{D-2}
\right)\nonumber \\
&\times &
\exp\left[\sum_{i=1}^{n}N_ih_i^{(s)} x^{i}
-2(\gamma-\gamma_{0})\right].
\end{eqnarray}
Here
\begin{equation}
G_{ij}=N_{i}\delta_{ij}-N_{i}N_{j}
\end{equation}
are the components of the minisuperspace metric,
\begin{equation}
V=
\kappa^2\sum_{s=1}^{m}A^{(s)}
\exp\left[\sum_{i=1}^{n}N_ih_i^{(s)} x^{i}
-2(\gamma-\gamma_0)\right].
\end{equation}
(2.10) is used to replace the densities $\rho^{(s)}$ in (2.11), (2.12)
by expressions of the functions $x^i(t)$.

After the gauge fixing
$\gamma=F(x^1, \ldots, x^n)$ the equations of motion (2.12) are
the Lagrange-Euler equations obtained from the Lagrangian
\begin{equation}
L={\rm e}^{\gamma_0-\gamma}\left(
\frac{1}{2}\sum_{i,j=1}^{n}G_{ij}\dot{x}^{i}\dot{x}^{j}-V
\right)
\end{equation}
under the zero-energy constraint (2.12).

Now we introduce an n-dimensional real vector space $\bf{R}^n$.
By $e_{1},\ldots e_{n}$ we denote the canonical basis in  $\bf{R}^n$
($e_{1}=(1,0,\ldots,0$) etc.). Hereafter we use the following vectors:

\noindent
1. the vector $x$ with components being the solution of the equations of
motion
\begin{equation}
x=x^1(t)e_1 + \ldots + x^n(t)e_n,
\end{equation}

\noindent
2. m vectors $u_s$, each of them one component of the
perfect fluid
\begin{equation}\label{2.17}
u_{s}=\sum_{i=1}^n\left(
h^{(s)}_{i} - \frac{\sum_{k=1}^{n}N_{k}h^{(s)}_{k}}{D-2}
\right)e_i.
\end{equation}
Let $<.,.>$ be a symmetrical bilinear form defined on $\bf{R}^n$
such that
\begin{equation}
<e_{i},e_{j}>=G_{ij}.
\end{equation}
The form is nongenerated and the inverse matrix to
$(G_{ij})$ has the components
\begin{equation}
G^{ij} = \frac{\delta^{ij}}{N_{i}}+\frac{1}{2-D}.
\end{equation}
The form  $<.,.>$ endows the space $\bf{R}^n$ with a metric, the signature of
which is
$(-, +, ..., +)$ \cite {I89},\cite {I89a}. $G_{ij}$ is used to
introduce the covariant components of vectors $u_{s}$
\begin{equation}
u^{(s)}_i=\sum_{i=1}^n G_{ij}u_{(s)}^j=N_ih^{(s)}_{i}.
\end{equation}
Form them the bilinear form reads
\begin{equation}
<u_{s},u_{r}>=
\sum_{i=1}^{n}h_{i}^{(s)}
h_{i}^{(r)} N_{i}+
\frac{1}{2-D}\left[\sum_{i=1}^{n}h_{i}^{(s)} N_{i}\right]
\left[\sum_{j=1}^{n}h_{j}^{(r)} N_{j}\right].
\end{equation}

A vector $y\in \bf{R}^n$ is called time-like, space-like
or isotropic, if $<y,y>$  is smaller,  greater  than or  equal to zero,
correspondingly. The vectors $y$ and $z$ are called orthogonal if
$<y,z>=0$.

Using the notation $<.,.>$ and the vectors (2.16)-(2.17), we may write the
zero-energy constraint (2.11) and the Lagrangian (2.15) in the form
\begin{eqnarray}%
E=\frac{1}{2}<\dot{x},\dot{x}>+
\kappa^2{\rm e}^{2(\gamma-\gamma_0)}
\sum_{s=1}^{m}A^{(s)}
{\rm e}^{<u_{s}, x>}
=0 ,
\label{E}
\\
L=\frac{1}{2}{\rm e}^{\gamma_0-\gamma}<\dot{x},\dot{x}>
-\kappa^2{\rm e}^{\gamma-\gamma_0}
\sum_{s=1}^{m}A^{(s)}
{\rm e}^{<u_{s}, x>} .
\label{L}
\end{eqnarray}

Furthermore, we take the so called harmonic time
gauge, which implies
\begin{equation} %
\gamma(t)=\gamma_0=\sum_{i=1}^{n} N_{i}x^{i}.
\end{equation}

>From the mathematical point of view the problem consist in solving the
dynamical system, described by a Lagrangian of the general form
\begin{equation}%
L=
\frac{1}{2}<\dot{x},\dot{x}>- \sum_{s=1}^ma^{(s)}
{\rm e}^{<u_{s}, x>},
\end{equation}
where $x,u_{s}\in \bf{R}^n$. It should be noted that the kinetic term
$<\dot{x},\dot{x}>$ is not a positively definite bilinear form as it
is usually the case
in classical mechanics. Due to the pseudo-Euclidean signature
$(-, +, ..., +)$ of the form $<.,.>$ such systems may be called
pseudo-Euclidean Toda-like systems as the potential given in
(2.25) defines a well known in classical mechanics Toda lattices \cite{Toda}.

Note that,
we have to integrate
the equations of motion following from the Lagrangian (2.25) under the
zero-energy constraint. Although an additional constant term
$-a^{(0)}$ (with $u_{0}\equiv 0\in \bf{R}^n$)
in the Lagrangian (2.25)
does not change the equations of motion, it nevertheless shifts
the energy constraint from zero to
\begin{equation}
E\equiv\frac{1}{2}<\dot{x},\dot{x}>+
\sum_{s=1}^{m}a^{(s)}\exp[<u_{s},x>]
=-a^{(0)} \equiv -\kappa^2 A^{(0)}.
\end{equation}
In our cosmological model, with \rf{2.8} and \rf{2.17},
such a term corresponds to a perfect fluid
with $h_{i}^{(0)}=0$ for all $i=1,\ldots ,n$.
This is in fact just a Zeldovich (stiff) matter component,
which can also be interpreted as a minimally  coupled
real scalar field.
Taking into account the possible presence of Zeldovich matter,
we have now to integrate the equations of motion
for an arbitrary energy level $E$.

\section{Solving the equations of motion for the model \penalty5000 \break
which reduces to a classical open Toda chain}
\setcounter{equation}{0}
We start from the Lagrangian (2.25) and the energy constraint (2.26) with
\begin{equation}
\label{3.1}
n\geq m+1\quad ,\quad m\geq 2\quad .
\end{equation}
The vectors $u_s$ may obey the relations
\begin{eqnarray}
\label{3.2}
<u_s,u_s>&=&u^2>0\quad ,\quad s=1,\ldots ,m\quad ,\\
\label{3.3}
<u_r,u_{r+1}>&=&-\frac{1}{2}u^2\quad ,\quad , r=1,\ldots ,m-1\quad ,\\
\label{3.4}
\mbox{all the remaining} \quad <u_r,u_s>&=&0\quad ,
\end{eqnarray}
where $u$ is an arbitrary non-zero real number. In this case the vectors
$u_s$ are space-like, linearly independent, and can be interpreted as root
vectors of the Lie algebra \linebreak $A_m=sl(m+1,\bf{C})$.
The Cartan matrix $\left(K_{rs}\right)$
(see e.g. \cite{Helgason,Humphreys})
then reads
\begin{equation}
\label{3.5}
\left(K_{rs}\right)=\left(\frac{2<u_r,u_s>}{<u_r,u_r>}\right)=\left(
\begin{array}{*{6}{c}}
2&-1&0&\ldots&0&0\\
-1&2&-1&\ldots&0&0\\
0&-1&2&\ldots&0&0\\
\multicolumn{6}{c}{\dotfill}\\
0&0&0&\ldots&2&-1\\
0&0&0&\ldots&-1&2
\end{array}
\right)\quad .
\end{equation}
Now, we choose in $\bf{R}^n$ a basis $\{f_1,\ldots,f_n\}$ with the following
properties
\begin{eqnarray}
\label{3.6}
f_{s+1}&=&\frac{2u_s}{<u_s,u_s>}\quad , \quad s=1,\ldots,m\quad ,\\
\label{3.7}
<f_1,f_i>&=&\eta_{1i}\quad ,\quad i=1,\ldots,n\quad ,\\
\label{3.8}
<f_{s+1},f_k>&=&0 \; ,\; <f_k,f_l>=\eta_{kl}\quad ,\quad
s=1,\ldots,m\quad ;\quad k,l=m+2,\ldots,n
\end{eqnarray}
with
\begin{equation}
\label{3.9}
\left(\eta_{ij}\right)=diag(-1,+1,\ldots,+1)\quad , \quad i,j=1,\ldots,n
\quad .
\end{equation}
Note that, the basis contains,
besides vectors proportional to $u_s$,
additional vectors $f_{m+2}$,\ldots,$f_n$, iff $n>m+1$.
By the decomposition
\begin{equation}
\label{3.10}
x(t)=\sum_{i=1}^n q^i(t)f_i\quad
\end{equation}
w.r.t. this basis,
with relations (\ref{3.2}) - (\ref{3.4}),
(\ref{3.6}) - (\ref{3.8}) the Lagrangian (2.25)
takes the form
\begin{eqnarray}
\label{3.11}
L&=&\frac{1}{2}\left(-\left(\dot{q}^1\right)^2+\frac{4}{u^2}
\left[\sum_{s=2}^{m+1}\left(\dot{q}^s\right)^2-
\sum_{p=2}^m\dot{q}^p\dot{q}^{p+1}\right]
+\sum_{k=m+2}^n\left(\dot{q}^k\right)^2\right)\nonumber\\
&&-a^{(1)}e^{2q^2-q^3}-\sum_{r=3}^ma^{(r-1)}e^{2q^r-q^{r-1}-q^{r+1}}
-a^{(m)}e^{2q^{m+1}-q^m}\quad .
\end{eqnarray}
The equations of motion for $q^1(t)$, $q^{m+2}(t)$,\ldots,$q^n(t)$ read
\begin{equation}
\label{3.12}
\ddot{q}^1(t)=0\;,\; \ddot{q}^{m+2}=0\; ,\; \ldots ,\; \ddot{q}^n(t)=0\; .
\end{equation}
Then,
\begin{eqnarray}
\label{3.13}
q^1(t)&=&a^1t+b^1\\
\label{3.14}
q^k(t)&=&a^kt+b^k\; ,\; k=m+2,\ldots,n.
\end{eqnarray}
The other equations of motion for $q^2(t)$,\ldots,$q^{m+1}(t)$ follow from
the Lagrangian
\begin{equation}
\label{3.15}
L_E=\sum_{s=2}^{m+1}\left(\dot{q}^s\right)^2
-\sum_{p=2}^m\dot{q}^p\dot{q}^{p+1}
-\frac{u^2}{2}\left[a^{(1)}e^{2q^2-q^3}
+\sum_{r=3}^m a^{(r-1)}e^{2q^r-q^{r-1}-q^{r+1}}
+a^{(m)}e^{2q^{m+1}-q^m}\right]\; .
\end{equation}
The linear transformation
\begin{equation}
\label{3.16}
q^{s+1}\longrightarrow q^s-\ln C^s\; ,\; s=1,\ldots,m\;,
\end{equation}
where the constants $C^1$,\ldots,$C^m$ have to satisfy
\begin{equation}
\label{3.17}
\sum_{s=1}^m K_{rs}\ln C^s=\ln \frac{u^2a^{(r)}}{2},\ r=1,\ldots,m\; ,
\end{equation}
brings the Lagrangian into the form
\begin{equation}
\label{3.18}
L_{A_m}=\sum_{s=1}^m \left(\dot{q}^s\right)^2
-\sum_{r=1}^{m-1} \dot{q}^r\dot{q}^{r+1}
-e^{2q^1-q^2}
-\sum_{p=2}^{m-1} e^{2q^p-q^{p-1}-q^{p+1}}
-e^{2q^m-q^{m-1}}\; .
\end{equation}
The latter represents the Lagrangian of a Toda chain associated with the Lie
algebra $A_m$ \cite{Toda} when the root vectors are put into the
Chevalley basis and coordinates describing the motion of the mass center are
separated out.

We use the method suggested in \cite{Anderson} for solving the equations of
motion following from (\ref{3.18}) and obtain
\begin{equation}
\label{3.19}
e^{-q^s}\equiv F_s(t)=\sum_{r_1<\ldots<r_s}^{m+1} v_{r_1}\cdots v_{r_s}
\Delta^2(r_1,\ldots,r_s)e^{(w_{r_1}+\ldots +w_{r_s})t}
\end{equation}
where $\Delta^2(r_1,\ldots,r_s)$ denotes the square of the Vandermonde
determinant
\begin{equation}
\label{3.20}
\Delta^2(r_1,\ldots,r_s)=\prod_{r_i<r_j}\left(w_{r_i}-w_{r_j}\right)^2\; .
\end{equation}
The constants $v_r$ and $w_r$ have to satisfy the relations
\begin{equation}
\label{3.21}
\prod_{r=1}^{m+1} v_r=\Delta^{-2}(1,\ldots,m+1)\; ,
\end{equation}
\begin{equation}
\label{3.22}
\sum_{r=1}^{m+1}w_r=0\; .
\end{equation}
The energy of the Toda chain described by this solution is given by
\begin{equation}
\label{3.23}
E_0=\frac{1}{2}\sum_{r=1}^{m+1}w^2_r\; .
\end{equation}
Finally, we obtain the folowing decomposition of the vector $x(t)$
\begin{equation}
\label{3.24}
x(t)=(a^1t+b^1)f_1+\sum_{s=1}^m \frac{-2\left(\ln F_s(t)+\ln
C^s\right)}{<u_s,u_s>}u_s
+\sum_{k=m+2}^m (a^kt+b^k)f_k\;
\end{equation}
We remind the reader that the coordinates $x^i(t)$ of the vector $x(t)$ are,
with respect to the {\it canonical} basis in $\bf{R}^n$, the logarithms of
the scale factors in the corresponding cosmological model.

Let us introduce the vectors
\begin{equation}
\label{3.25}
\alpha =a^1f_1 +\sum_{k=m+2}^n a^kf_k\equiv\sum_{i=1}^n\alpha^ie_i\in
{\bf{R}}^n
\end{equation}
\begin{equation}
\label{3.26}
\beta =b^1f_1 +\sum_{k=m+2}^n b^kf_k\equiv\sum_{i=1}^n\beta^ie_i\in
{\bf{R}}^n
\end{equation}
with $\alpha^i$, $\beta^i$ being their coordinates with respect to the
canonical basis. Using (\ref{3.7}) and (\ref{3.8}), we conclude these
coordinates have to satisfy the following equations
\begin{equation}
\label{3.27}
<\alpha,u_s>=\sum_{i,j=1}^n G_{ij}\alpha^iu_{(s)}^j=0\; ,\;s=1,\ldots,m\; ,
\end{equation}
\begin{equation}
\label{3.28}
<\beta,u_s>=\sum_{i,j=1}^n G_{ij}\beta^iu_{(s)}^j=0\; ,\;s=1,\ldots,m\; ,
\end{equation}
where the $u_{(s)}^i$ are the coordinates of $u_s$ in the canonical basis
(see (2.17)).

The total energy $E$ of the system is given by
\begin{equation}
\label{3.29}
E=\frac{1}{2}<\alpha,\alpha>+\frac{2}{u^2}E_0=\frac{1}{2}\sum_{i,j=1}^n
G_{ij}\alpha^i
\alpha^j+\frac{1}{u^2}\sum_{s=1}^{m+1} \left(w_s\right)^2\; .
\end{equation}
If $m=n+1$, then $<\alpha,\alpha>=-\left(a^1\right)^2\le 0$.
With (\ref{3.29}), we then obtain $E\le \frac{2}{u^2}E_0$.

Finally, the scale factors of the multidimensional cosmological model
with the Lagrangian (2.25) and the energy constraint (2.26) are given by
\begin{equation}
\label{3.30}
e^{x^i(t)}=\prod_{s=1}^m \left[\tilde{F}^2_s(t)\right]^{-u_{(s)}^i/<u_s,u_s>}
e^{\alpha ^it+\beta ^i}\quad ,
\end{equation}
where
\begin{equation}
\label{3.31}
\tilde{F}_s(t)=C^s \cdot F_s(t)\; ,\; s=1,\ldots,m\; .
\end{equation}
Using (2.10) we obtain the following solution for the densities of the
perfect
fluid components
\bear{3.32}
\rho^{(1)}=A^{(1)}{\rm e}^{-2\gamma_0}
\frac{\tilde{F}_{2}}{\tilde{F}^{2}_1}
\ ,\quad
\rho^{(m)}=A^{(m)}{\rm e}^{-2\gamma_0}
\frac{\tilde{F}_{m-1}}{\tilde{F}^{2}_m}
\nn\\
\rho^{(p)}=A^{(p)}{\rm e}^{-2\gamma_0}
\frac{\tilde{F}_{p-1}\tilde{F}_{p+1}}{\tilde{F}^{2}_p}
\ ,\quad p=2,\ldots,m-1.
\ear
where $\gamma_0$ is defined by (2.5) and may be calculated by (3.30).

The constants $C^s$ are specified by (3.17). The solution contains the
parameters $\alpha^i,\ \beta^i,\ v_r,\ w_r$ ($i=1,\ldots,n,\
r=1,\ldots,m+1$) obeying the constraints (3.27),(3.28),(3.21),(3.22),(3.29).
If the energy $E$ is arbitrary
(see (2.26))
the solution has $2n$ free parameters as
required.

\section{Example}
\setcounter{equation}{0}
We consider a space-time manifold $M$ with the following structure
\begin{equation}
\label{4.1}
M=R\times M_1^3\times M_2^3\times M_3^4
\end{equation}
where the dimensions of $M_1^3$, $M_2^3$, and $M_3^4$ are $3$, $3$,
and $4$,
respectively.
The first component of the perfect fluid shall have the $h_i^{(1)}$ values
\begin{equation}
\label{4.2}
h_1^{(1)}=0\quad, \quad h_2^{(1)}=h\quad ,\quad h_3^{(1)}=0
\end{equation}
while the second component is given by
\begin{equation}
\label{4.3}
h_1^{(2)}=h\quad, \quad h_2^{(2)}=0\quad ,\quad h_3^{(2)}=0\quad .
\end{equation}
$h$ is a real valued parameter with the restriction
\begin{equation}
\label{4.4}
h \neq 0 .
\end{equation}
It is easy to check that the relations \rf{3.2}, \rf{3.3} with $s=2$ and
$r=1$ are fulfilled.

In this case, the exact solution of the field equations gives the metric
\begin{eqnarray}
\label{4.5}
ds^2&=&\left[\tilde{F}_1(t)\tilde{F}_2(t)\right]^{\frac{2}{3h}}\nonumber\\
&&\!\!\!\!\!\!\!\!\!\!\!\! \times \Bigg\{
-\exp(8\alpha_0t+8\beta_0)dt^2
+\frac{ds_1^2}{\tilde{F}_2^{\frac{2}{h}}(t)}
+\frac{ds_2^2}{\tilde{F}_1^{\frac{2}{h}}(t)}+
\exp(2\alpha_0t+2\beta_0)ds^2_3\Bigg\}
\end{eqnarray}
with the following definitions
\begin{equation}
\label{4.6}
\tilde{F}_1(t)=\kappa^2\left[A^{(1)}\right]^{\frac{2}{3}}
\left[A^{(2)}\right]^{\frac{1}{3}}h^2F_1(t)\quad ,
\end{equation}
and
\begin{equation}
\label{4.7}
\tilde{F}_2(t)=\kappa^2\left[A^{(1)}\right]^{\frac{1}{3}}
\left[A^{(2)}\right]^{\frac{2}{3}}h^2F_2(t)\quad ,
\end{equation}
and
\begin{equation}
\label{4.8}
F_1(t)=v_1e^{w_1t}+v_2e^{w_2t}+v_3e^{w_3t}\quad ,
\end{equation}
\begin{eqnarray}
\label{4.9}
F_2(t)&=&v_1v_2(w_1-w_2)^2e^{(w_1+w_2)t}+v_1v_3(w_1-w_3)^2e^{(w_1+w_3)t}
\nonumber\\
&&+v_2v_3(w_2-w_3)^2e^{(w_2+w_3)t}\quad .
\end{eqnarray}
In our case, the energy $E$ is given by
\begin{equation}
\label{4.10}
E=-6\alpha^2_0+\frac{1}{2h^2}
\left[w_1^2+w_2^2+w_3^2\right]=-\kappa^2A^{(0)}
\quad .
\end{equation}
$A^{(0)}>0$ means that Zeldovich matter is present in all the internal
spaces (See the remarks at the end of sect.~3).
With $A^{(0)}=0$ \rf{4.10}
is the energy constraint specialized to our example.

The nine parameters $w_1, w_2, w_3, v_1, v_2, v_3, \alpha_0, \beta_0, E$
have to satisfy \rf{4.9} and the two further relations
\begin{equation}
\label{4.11}
w_1+w_2+w_3=0
\end{equation}
and
\begin{equation}
\label{4.12}
v_1v_2v_3=\left[(w_1-w_2)(w_2-w_3)(w_1-w_3)\right]^{-2}
\end{equation}
(See \rf{3.22} and \rf{3.21}!).

Finally, we have to give the expressions for the matter densities
$\rho^{(1)}$ and $\rho^{(2)}$. They read
\begin{equation}
\label{4.13}
\rho^{(1)}=A^{(1)}\left[\tilde{F}_1^{-2/(3h)-2}(t)\tilde{F}_2^{1-2/(3h)}(t)
\right]
e^{-8\alpha_0t-8\beta_0}\quad ,
\end{equation}
\begin{equation}
\label{4.14}
\rho^{(2)}=A^{(2)}\left[\tilde{F}_2^{-2/(3h)-2}(t)\tilde{F}_1^{1-2/(3h)}(t)
\right]
e^{-8\alpha_0t-8\beta_0}
\end{equation}
and their quotient is
\begin{equation}
\label{4.15}
\frac{\rho^{(2)}}{\rho^{(1)}}=\frac{A^{(2)}}{A^{(1)}}
\frac{\tilde{F}_1(t)}{\tilde{F}_2(t)} \quad .
\end{equation}
Although the solution is invariant with respect to the exchange $(w_1, w_2,
v_1, v_2) \rightarrow (w_2, w_1, v_2, v_1)$ there is still enough freedom
to build a lot of solutions that it is difficult to recognize many
general properties
of the solutions. What one can say is the following: We know that
\begin{equation}
\label{4.16}
\left(e^{x^1(t)}\right)^{3h}\propto\left|\frac{F_1(t)}{F_2^2(t)}\right|
\end{equation}
and
\begin{equation}
\label{4.17}
\left(e^{x^2(t)}\right)^{3h}\propto\left|\frac{F_2(t)}{F_1^2(t)}\right|
\quad  .
\end{equation}

\unitlength1cm
\begin{figure}[tb]
\begin{picture}(0,0)(-5.1,4)   
\put(0,0){$e^{x^3_1}$}
\end{picture}
\begin{picture}(0,0)(-8.4,3.5)   
\put(0,0){$e^{x^1_1}$}
\end{picture}
\begin{picture}(0,0)(-7.9,9.3)   
\put(0,0){$e^{x^1_2}$}
\end{picture}
\begin{picture}(0,0)(-9.7,9.2)   
\put(0,0){$e^{x^2_1}$}
\end{picture}
\begin{picture}(0,0)(-7.9,11)    
\put(0,0){$e^{x^2_2}$}
\end{picture}
\begin{picture}(0,0)(-10.8,11)   
\put(0,0){$e^{x^3_2}$}
\end{picture}
\begin{center}
\leavevmode
\epsfxsize=300bp \epsfbox{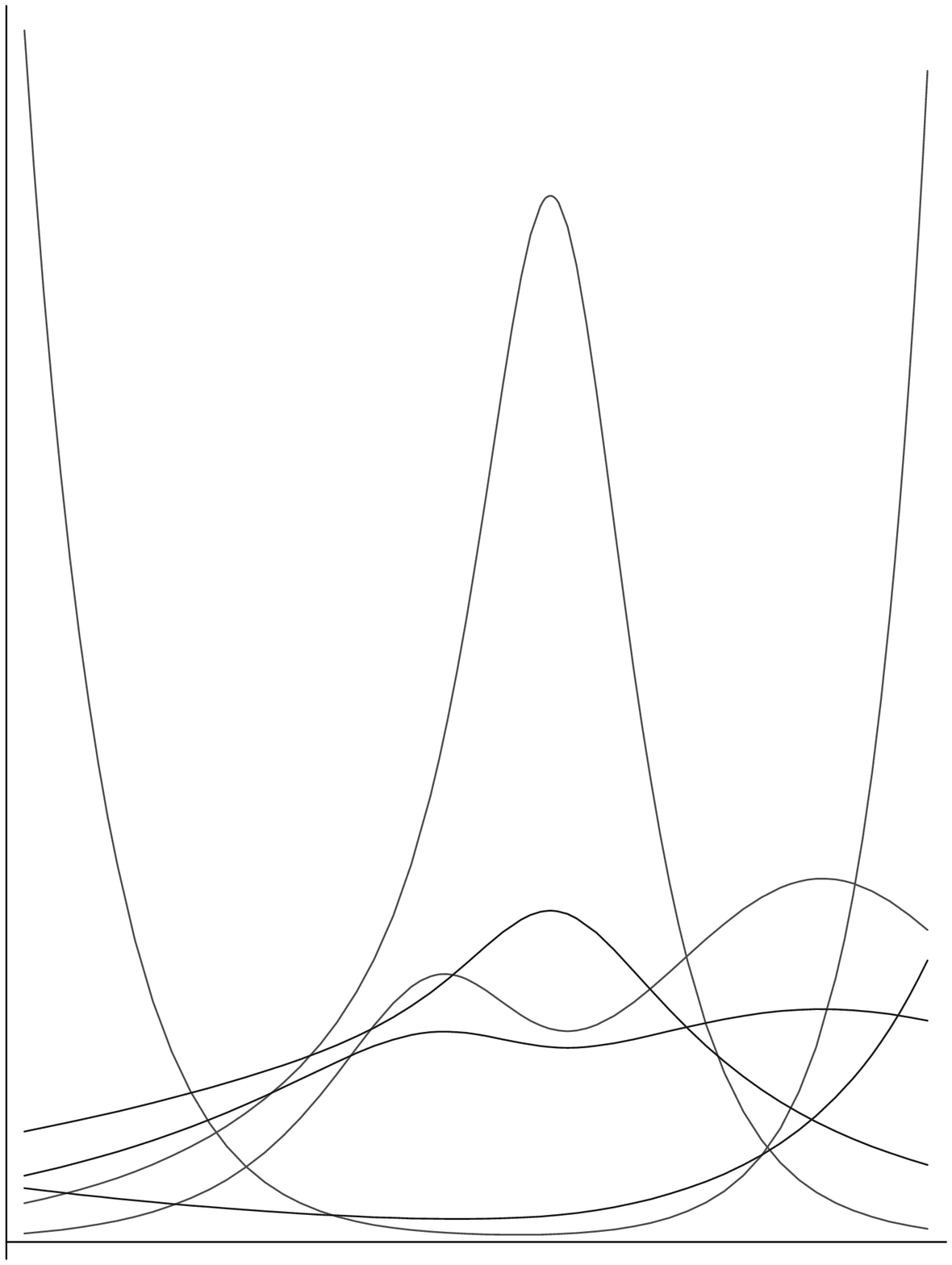}
\end{center}
For the parameter values $v_1=1$, $v_2=10$, $w_1=1.1$, $w_2=5$
and two different values of $h$ ($h_1=2/3$, $h_2=2$) (Zeldovich matter is
present) the scale factors $x^i(t)$
are pictured on a certain time intervall. $x^i_k$ is $x^i$ for $h=h_k$.
The integration constants $A^{(s)}$
are put equal to $1$.
\end{figure}

\unitlength1cm
\begin{figure}[tb]
\begin{picture}(0,0)(-11.1,2.3)   
\put(0,0){$\rho^1_1$}
\end{picture}
\begin{picture}(0,0)(-10.9,10.3)  
\put(0,0){$\rho^2_1$}
\end{picture}
\begin{picture}(0,0)(-4.6,2.8)    
\put(0,0){$\rho^1_2$}
\end{picture}
\begin{picture}(0,0)(-9.4,11.7)     
\put(0,0){$\rho^2_2$}
\end{picture}
\begin{center}
\leavevmode
\epsfxsize300bp \epsfbox{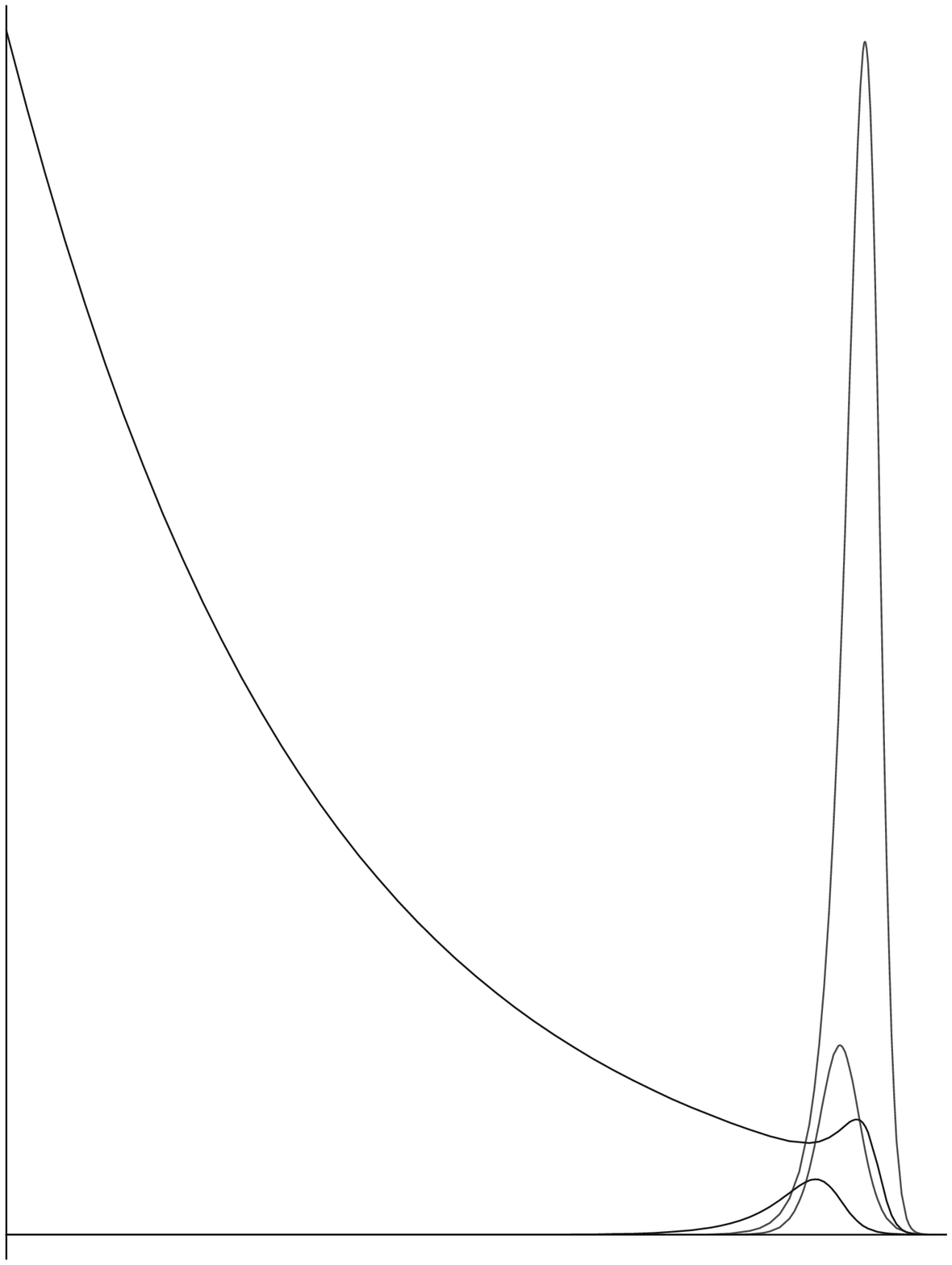}
\end{center}
The energy densities $\rho_1$ and $\rho_2$ belonging to
the parameter values $v_1=1$, $v_2=10$, $w_1=1.1$, $w_2=5$
and two different values of $h$ ($h_1=2/3$, $h_2=2$) (Zeldovich matter is
present). $\rho_i^k$ means the energy density
$\rho_i$ belonging to $h_k$.
\end{figure}

An easy but tedious discussion of the different possibilities of choosing
the
parameters $w_1$ and $w_2$ shows that the expressions \rf{4.16} and
\rf{4.17} have for $t\rightarrow \pm \infty$ the following shape:
\begin{equation}
\label{4.18}
\left|\frac{F_2(t)}{F_1^2(t)}\right|\quad
\stackrel{t\rightarrow \pm \infty}{\longrightarrow}\quad e^{f_{\pm
\infty}(w_1,w_2)t}
\end{equation}
and
\begin{equation}
\label{4.19}
\left|\frac{F_1(t)}{F_2^2(t)}\right|\quad
\stackrel{t\rightarrow \pm \infty}{\longrightarrow}\quad e^{g_{\pm
\infty}(w_1,w_2)t}
\end{equation}
with some functions $f_{\pm \infty}$, and $g_{\pm \infty}$ of the parameters
$w_1$ and $w_2$ being negative for $+\infty$ and positive for $-\infty$.
This shows that the scale factors of the manifold $M_1$ and $M_2$ go to zero
for $t\rightarrow \pm \infty$.Because of the behaviour of the scale factor
of $M_1$,
as one could have
expected, the solution can not properly describe the total cosmic evolution.
But what one can do is arranging especially the parameters $w_1$ and $w_2$
such that the behaviour of the scale factors is acceptable from a physical
point of view (see below).

As for the scale factor $e^{x^3(t)}$ of the manifold $M_3$, we have
\begin{equation}
\label{4.20}
\left(e^{x^3(t)}\right)^{3h}\propto \left|F_1F_2\right|e^{3\alpha_0ht}
\end{equation}
(the parameter $\beta_0$ shall be absorbed by coordinates). For
$t\rightarrow \pm \infty$ we have
\begin{equation}
\left|F_1F_2\right|\quad \rightarrow \quad e^{\pm k(w_1, w_2)t}
\end{equation}
where $k(w_1, w_2)$ is some positive function of the parameters $w_1$ and
$w_2$.

The proper time $T$
as a function of harmonic time $t$ is
given by integration of $dT=e^{\gamma_0}dt$ with
$\gamma_0 = 3x^1+3x^2+4x^3$.
For $t\rightarrow \pm \infty$ the behaviour of  both, the proper
time
$T$
and
the scale factor $e^{x^3}$
depends on the choice of $\alpha_0$.
This holds for the case that in all manifolds
Zeldovich matter is present.
If the case is excluded then $\alpha_0$ is given
by the energy constraint.
We shall not discuss this point here because of
the above mentioned  restricted range of possibly physical importance of the
solution. Instead, we present some drawings conveying how the scale factors
and matter densities can behave for special values of $w_1$, $w_2$, $v_1$,
$v_2$, and $\alpha_0$.


\begin{center}
{\bf Acknowledgments}
\end{center}
\par
The authors are grateful for hospitality at Astrophysikalisches Institut Potsdam.
This work was supported by DFG grants  436 RUS 113/236, Kl 732/4-1 and Schm 911/6.

\pagebreak

\end{document}